\documentclass[5p]{elsarticle}

\usepackage[hidelinks]{hyperref}

\journal{Journal of \LaTeX\ Templates}

\usepackage{listings}
\usepackage{multirow}
\usepackage{rotating}
\usepackage{longtable}
\usepackage{multicol}
\usepackage{lipsum}
\usepackage{booktabs}
\usepackage{xcolor}
\usepackage{colortbl}
\usepackage{rotating}
\usepackage{multirow}
\usepackage{booktabs}
\usepackage{array}
\usepackage{tikz}
\usepackage{enumitem}
\def\checkmark{\tikz\fill[scale=0.4](0,.35) -- (.25,0) -- (1,.7) -- (.25,.15) -- cycle;} 

\definecolor{blue}{rgb}{0.0, 0.0, 255}
\definecolor{red}{rgb}{255, 0.0, 0.0}

\begin{document}

\begin{frontmatter}


\title{Semantics-based Privacy by Design for Internet of Things Applications}

\author[add1]{Lamya Alkhariji }
\fntext[myfootnote]{Corresponding author at: AlkharijiLa@cardiff.ac.uk. Sponsored by  Imam Muhammad bin Saud University}

\author[add2]{Suparna De }

\author[add1]{Omer Rana }

\author[add1]{Charith Perera}

\address[add1]{Cardiff University, Cardiff CF10 3AT, United Kingdom}
\address[add2]{University of Surrey, Guildford GU2 7XH, United Kingdom}



\begin{abstract}
As Internet of Things (IoT) technologies become more widespread in everyday life, privacy issues are becoming more prominent. The aim of this research is to develop a personal assistant that can answer software engineers' questions about Privacy by Design (PbD) practices during the design phase of IoT system development. Semantic web technologies are used to model the knowledge underlying PbD measurements, their intersections with privacy patterns, IoT system requirements and the privacy patterns that should be applied across IoT systems. This is achieved through the development of the PARROT ontology, developed through a set of representative IoT use cases relevant for software developers. This was supported by gathering Competency Questions (CQs) through a series of workshops, resulting in 81 curated CQs. These CQs were then recorded as SPARQL queries, and the developed ontology was evaluated using the Common Pitfalls model with the help of the Protégé HermiT Reasoner and the Ontology Pitfall Scanner (OOPS!), as well as evaluation by external experts. The ontology was assessed within a user study that identified that the PARROT ontology can answer up to 58\% of privacy-related questions from software engineers. 
\end{abstract}


\end{frontmatter}


\section{Introduction}
There have recently been significant increases in the deployment of Internet of Things (IoT) systems, extending into domains as varied as smart homes, personal health, wearables and public space monitoring, many of which entail the collection and manipulation of large quantities of user data \cite{De2021} \cite{LAROUI2021210}. Protecting privacy of an individual within such systems has become a growing concern that requires urgent attention. In recent years, there has been increasing interest in the development and enforcement of privacy laws and practices by various authorities. One main issue for many software engineers is that these rules are often complex and abstract in nature, and thus require field experts to translate them into more implementable formats. While this is feasible for large companies, in small-to-medium enterprises (SMEs), such bespoke approaches can become an unbearable burden, and consequently are often neglected. 

Privacy by Design (PbD) is a concept that suggests considering data protection during the system design phase, leading to a more practical solution to satisfy a data subject's\footnote{An individual using the IoT system whose data are collected} privacy \cite{Wong2019}. In this context, several PbD measurements (or schemes) at different levels of abstraction, including principles, guidelines, strategies and privacy patterns, have been proposed by various organisations and researchers, as reviewed in~\cite{Alkhariji2020}. However, the software developer must still determine which PbD practices are best suited to the system under development, which nevertheless adds significant effort to the development process. This paper introduces a solution that enables software engineers to query the components of the system under development with respect to the privacy measurements required. 

Semantic web technologies enable structured annotation, integration and retrieval of massive quantities of data \cite{De2017}. In this paper, we introduce the PARROT ontology, which models IoT system needs and binds them to the relevant PbD measurements. The contributions of this paper are as follows:

\begin{enumerate}[leftmargin=1em]

\item An analysis of PbD needs in IoT systems, obtained from actual software engineers in the form of Competency Questions (CQs)\footnote{Question expressed in natural language by stakeholders that defines the scope of the ontology} within real IoT use cases and their corresponding privacy patterns. This should enable researchers, privacy professionals, and standards organisations to achieve better design for privacy protection.

\item The introduction of the PARROT ontology, which encapsulates existing PbD measurements and their inter-relationships as a means of offering easily explainable PbD guidance. In addition, the PARROT ontology captures the knowledge required to answer software engineers’ questions on privacy when designing IoT systems.

\item An assessment of the PARROT ontology quality across three different aspects, and a demonstration of the use of the PARROT ontology within a user-based study. 

\end{enumerate}

The remainder of this paper is structured as the following: Section~2 presents a motivating scenario to provide a context for this paper. Section~3 presents related work covering existing privacy regulations and ontologies. Section~4 explains the methodology we followed in the research. Section~5 shows how we gathered requirements for the PARROT ontology. Section~6 provides an analysis of the information we gathered. Section~7 explores the PARROT ontology specifications and description. In Section~8 we validated the PARROT ontology using CQs. In Sections~9 and~10, we evaluate the implementation of the PARROT ontology. In Section~11 we discuss the results of the evaluation, concluding the paper in Section~12.



\section{Motivating Scenario}
To illustrate the use of the tool under development, a case study featuring "Nora", a software developer who seeks to implement GDPR rule of PbD in the system she is developing is considered.
\subsection{Scenario}
Nora is developing an IoT system, but she needs to think about user privacy. She searches about the required PbD practices in order to understand them and then determines which ones she needs to apply. She finds many resources and documents that describe ways to protect the privacy of users, but she is confused by the large number of available documents and their variations. For example, Cavoukian’s principle, ``Proactive, not Reactive; Preventative not Remedial” that she finds easy to understand, but these are vague in application. Looking at another document, Hoepman’s strategies, she is unsure whether she needs to apply all of the strategies. On looking at the first strategy, “Minimise”, she approaches her system with the intent of minimising data. This becomes somewhat confusing where she seeks further explanations of this strategy, discovering the privacy patterns document which she finds applicable. However, it offers so many patterns that she is not sure which ones best explain the “Minimise” strategy. She goes back to her system having struggles to find the appropriate practices to deploy.

Nora thus decides to use a personalized assistant tool. She draws the system she is designing in the tool’s interface, and once she submits the DFD diagram, the tool returns it with annotations and comments about the privacy patterns required for each node in the diagram. She explores these comments, which help her ascertain what she needs to do to implement the appropriate patterns. Seeking further explanation, she uses a chatbot to ask questions about the meaning of these privacy patterns.

\subsection{Comments and Discussion}
Nora can thus finish her task more easily, having gained sufficient awareness of the privacy measurements required in her system. The chatbot function gives her the ability to ask questions about why a particular privacy pattern is advised and the relevance of each privacy pattern to the other measurements involved. This personalisation of the privacy assistant tool thus helps Nora to use the most appropriate privacy measurements across her system design.

%

\section{Related Work}
\subsection{Privacy Regulations and Standards}
As user privacy has become a prominent concern, it has been further protected by various legislative bodies in many countries. In Europe and the UK, the General Data Protection Regulation\footnote{gdpr-info.eu}(GDPR) \cite{EuropeanUnion2018}  is applied, whereas in the United States, different federal laws and regulations have been implemented by various state governments, such as Californian Consumer Privacy Act\footnote{oag.ca.gov/privacy/ccpa} (CCPA) and the Stop Hacks and Improve Electronic Data Security Act\footnote{privacyshield.gov} (SHIELD). In Australia, the Australian Privacy Principles (APPs) are used as a privacy protection framework\footnote{oaic.gov.au/privacy/australian-privacy-principles}. Aljeraisy et al. \cite{aljeraisy2021privacy} offer a more comprehensive analysis of privacy protection laws across different countries.

Meeting the requirements of all of these various laws can present a challenge to software engineers, particularly because of the unfamiliar language used in describing these requirements. This leads to a need to transform these laws into software requirements, a process referred to as Privacy by Design (PbD) \cite{Funds2006} \cite{Wong2019}. There have been multiple PbD measurements deployed by different parties at various levels, such as the seven privacy principles published by Cavoukian \cite{cavoukian2012privacy} and the eight privacy strategies created by Hoepman \cite{Hoepman2018}. Moreover, Perera et al. \cite{Perera2016} published 30 privacy guidelines specifically related to IoT systems, though the technical report \cite{alkhariji2020examining} reviews 10 PbD measurements published by different organisations and researchers along with their relationships to each other as a way of broadening scope. These PbD measurements are the ground source of information that we are using in this research, where we recommend consistent ones to the system design nodes provided.


\subsection{Privacy Information Needs}
A limited number of studies have explored the awareness of PbD regimes among software engineers. Perera et al. \cite{Perera2017} undertook an observational study to show how the creation of assistive structured privacy guidelines could be helpful in allowing software engineers to improve data subject privacy within their systems. They found that, irrespective of engineers' level of expertise, such guidelines led to similar levels of incorporation of privacy practices in the resulting designs. In addition, the study made clear that providing software engineers with a privacy guideline list affects design success, with a success rate of 75.12\%. 

Providing personalised assistants for software engineers is likely to have an impact on compliance with PbD practices, which drive improvement of data subject privacy. This is particularly true, based on automated assistant systems' proven ability to support clients efficiently \cite{Gharib2017}.  

The current work has incorporated as many PbD measurements as possible to embrace the idea of “explainable privacy”. This has been done because, among the various different types of PbD measurements, \textit{privacy patterns} are the most suitable for implementation by software engineers, yet other scheme levels, being more abstract, offer better descriptions of the aims behind each practice. The next section thus develops the concept that the knowledge underlying PbD can be translated into a machine-interpretable format in order to facilitate automation of PbD recommendations for IoT systems. 


\subsection{Ontologies Representing Domain Knowledge}
Ontologies, as a technology in the semantic web, offer reasonable means of representing a knowledge base. An ontology can thus represent a very wide range of concepts, along with their relationships and interactions, in a machine-readable format. For example, Dragoni et al. \cite{Dragoni2018} considered the adoption of advanced technology into an individual's lifestyle as a way to develop recommendations for personalised healthy practices by applying an ontology-centric decision support system called PerKApp. The proposed ontology provided expert knowledge and the information required to assist the user in developing healthy practices. It also incorporated semantic rules to act as expert support for the user’s healthy practices, identifying any violations in such practices, and notifying the user by means of motivational messages as required. That system was tested within the Key to Health project and thus found to be applicable in real-world scenarios. 

In another example, Malone et al. \cite{Malone2014} applied semantic technologies to achieve data reproducibility in the bioinformatics field. Their motivation was their belief that data analysis results vary depending on the software used for such analysis. To make data results more easily reproducible, researchers thus need to know the details of the software used to analyse the data. In ordered to build the required Software Ontology (SWO), they followed Agile methodology principles, as well as involving various types of participants as ontology users. The resulting SWO ontology was later merged with the EDAM \cite{ison2013edam} ontology, which was designed to handle bioinformatics operations, data types and identifiers, topics, and formats, and the resultant joint ontology was used in various biomedical applications, including the BioMedBridges software registry \cite{BioMedBridges2015}, eagle-I \cite{vasilevsky2012research}, and the Gene Expression Atlas Data project \cite{EMBL-EBI}. 
These examples illustrate ontologies as an effective technology to supply assistive systems. Hence, in this research, we are developing the PARROT ontology that fulfills our purpose. Many methodologies exist to guide ontology developers in creating associated analytical studies to compare options \cite{Rekha2017} \cite{Abdelghany2019} \cite{Kotis2020}. For this work, however, the NeON \cite{Gomez-Perez} and Chaware et al. \cite{Chaware2010} methodologies were adopted.

\subsection{Ontologies for Privacy by Design}
Multiple ontologies have been developed to support increased rigour in the privacy field. Harshvardhan et al. \cite{Pandit2018} attempted to address the complexity of understanding privacy policies by transforming such policies into machine-readable data. They proposed an ontology design pattern (ODP) that contains all details in a given privacy policy document, such as those on collection, usage, storage, and sharing of personal data, along with the relevant processes and legal basis in the GDPR. This ODP would thus have benefits above and beyond those of the GDPRov \cite{pandit2017modelling} and GDPRtEXT \cite{Pandit2018} ontologies, which cover the vocabulary, concepts, and terms within the GDPR. They designed an ODP to answer a set of competency questions related to personal data;  further competency questions about how personal data may be changed, deleted, and obtained were not incorporated at that stage. The authors thus acknowledged that the ODP required wider patterns to include all information in a privacy policy document in order to develop it into an ontology that could allow the full manipulation and understanding of the use of personal data. They modeled the information of privacy policies, whereas in our ontology we modeled PbD knowledge which are the structures to be followed in the design phase of software development.  

Gharib et al. \cite{Gharib2020} applied the PbD concept, rather than focusing only on security requirements,  as a solution to privacy breaches. However, they suggested that the vagueness of this privacy concept confuses designers and stakeholders, preventing them from making the right design decisions. To address this, they suggested that ontologies offer a more robust means of conceptualising privacy concepts and their interrelations, and to develop a relevant ontology, they systematically reviewed the literature to identify key concepts and relationships underlying general privacy requirements. From this review, they identified 38 key concepts and relationships, which they grouped across four categories, creating 17 organisational factors, nine risks, five treatments, and seven privacy factors. Although their concern is PbD requirements, their objective varies from the PARROT ontology that they aim to provide the software developer with generic privacy key concepts where we provide explainable PbD measurements that are matching and should be applied in the corresponding IoT system. 

\section{Methodology}
\begin{figure*}
	\centering
	\includegraphics[scale=0.5]{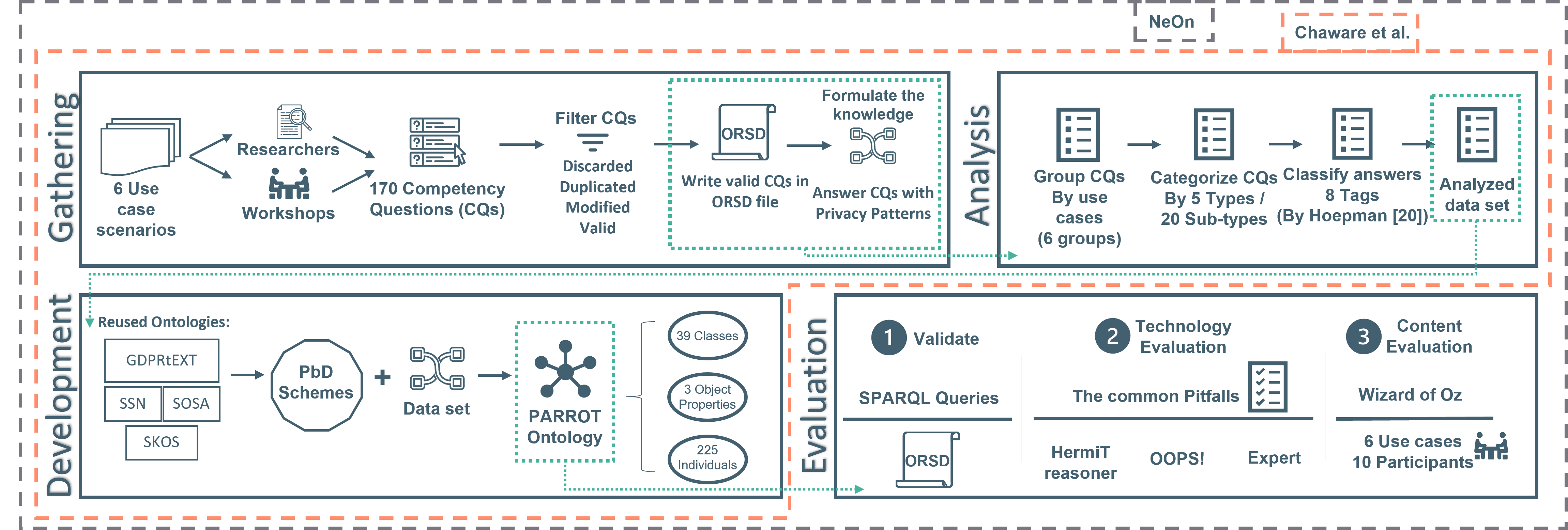}
	\caption{The figure shows the methodology followed in this work. There are four steps: gather, analyse, develop, and evaluate. The overall methodology of developing the PARROT ontology follows NeOn methodology \cite{Gomez-Perez}. That is integrated with Chaware et al. \cite{Chaware2010} methodology for the first three steps, i.e. gather, analyse, and develop.}
    \label{Figure:methodology}
\end{figure*}

The progress of the current research was organised into four phases as shown in figure \ref{Figure:methodology}. These were information gathering, analysis, development, and evaluation. The first step was to \textbf{gather} the information required for modelling in the PARROT ontology via six representative IoT use cases with different system components and data types. This step involved two sources of questions, those asked by researchers and by software engineers in a series of workshops. This resulted in the development of 170 competency questions (CQs) that were input to the filtration step. All resulting valid questions were then used to create an ontology requirements specification document (ORSD), which listed 81 CQs. For the final step in this phase, the answers to the retained CQs were determined and then formulated as a set of privacy patterns. At that point, the ORSD and the formulated knowledge were thus generated for the next phase.  Section \ref{Gathering PARROT Information Needs} discusses further details of the gathering phase.

In the \textbf{analyse} phase, the knowledge from the gathering phase was grouped, categorised, and tagged. The CQs were initially grouped depending on the use cases from which they were inferred; they were then categorised, depending on the issues raised, into five types and 20 sub-types. In addition, relevant answers were assigned to all CQs in the form of privacy patterns, using Hoepman's \cite{Hoepman2018} eight tags, as ascertained from previous research \cite{Alkhariji2020}. The necessary analysis to achieve this is discussed in further detail in section \ref{Analysis}. The analysed data sets were then moved to the next phase. 

In the \textbf{develop} phase, the PARROT ontology was created using a top-down approach. As a starting point, four existing ontologies, SKOS, GDPRtEXT, SSN, and SOS were reused, with classes created to model the knowledge that was to be included in the PARROT ontology. The PbD measurements and their connections were thus modelled together, based on previous work  \cite{Alkhariji2020}, along with the data set analysed in the previous phase. Further details of this development are offered in section \ref{PARROT ontology}.

Finally, the PARROT ontology was \textbf{evaluated} in three steps: 
(1) the CQs were validated in ORSD via SPARQL queries;
(2) the technology of the PARROT ontology was evaluated against the 41 pitfalls, as proposed by Villalón \cite{poveda2016ontology}, with this examination completed using three methods: 1. Application of the Protégé HermiT Reasoner; 2. Evaluation with the Ontology Pitfall Scanner (OOPS!); and 3. Lexical Semantic Expert Evaluation;
(3) the content of the PARROT ontology was then evaluated using the Wizard of Oz technique via a user study. 
The overall evaluation is thus discussed in sections \ref{Validation}, \ref{Technology Evaluation}, and \ref{Content Evaluation}. 
Across all four phases, the work was guided by the NeOn methodology \cite{Gomez-Perez}, a well known method of ontology development, as integrated with the methodology designed by Chaware et al. \cite{Chaware2010}, which provided supplementary practical steps for ontological data gathering and development for the first three phases. 

\section{Gathering PARROT Information Needs} \label{Gathering PARROT Information Needs}

To model the PARROT ontology, it was first necessary to identify the information required. A list of CQs that might be asked by a software developer seeking to apply privacy practices in a system was thus developed, with CQs extracted from six real different IoT use cases. The use cases cover a range of different contexts and purposes named: (1) Health care system, (2) Real-time tracking system, (3) Fitness watch, (4) Park monitoring system, (5) Smart home system, and (6) Drone delivery system.  We listed the use cases descriptions and diagrams and the CQs in \cite{orca149337}. The overall process of this created a data flow diagram (DFD) with an IoT system as input that then provides the software developer with a related list of privacy patterns matching the DFD components. The list of CQs was finalised in two stages: the initial set was drawn from the researchers' knowledge of IoT systems and Privacy Practices and as extracted from the IoT use cases noted above; then, further CQs were solicited from software engineers who were given various IoT use cases in a series of focused workshops. The answers to all CQs were then developed as a list of associated privacy patterns.

\subsection{Researcher-generated CQs}
The \textit{Health Care} use case was selected as an example of a system that collects sensitive information about data subjects. Healthcare applications are a growing IoT application domain, and they also are complex applications that often include multiple sensors and inputs, thus generating significant data volumes. As a result, they are ideal candidates for measurements aiming to preserve privacy, thus offering a good initial example for this work. The development of CQs was then undertaken in several steps: (1) A DFD for the selected use case was created; (2) A list of applicable privacy patterns was initiated; (3) The privacy patterns were manually allocated across the DFD; (4) CQs were then determined based on all nodes in the DFD. 
The following explains each step in detail:

\subsubsection{Health Care use case}

\begin{figure}
	\centering
	\includegraphics[scale=0.28]{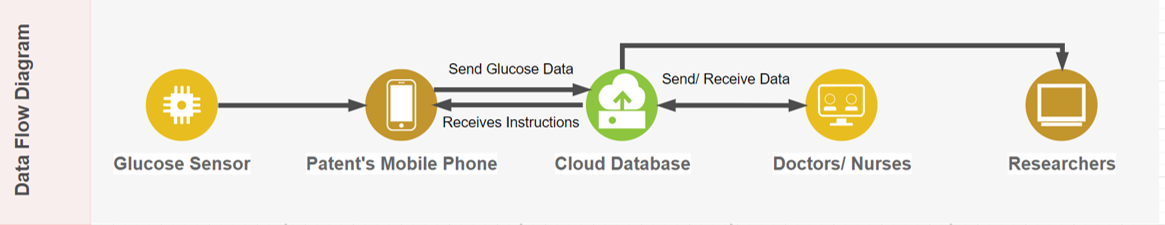}
	\caption{The DFD of Diabetes Treatment and Monitoring use case that was used to find the CQs by researchers. }
    \label{Figure:DiabetesTreatmentandMonitoringDFD}
\end{figure}

Health Care System is an IoT application that analyses patient health data in order to issue relevant alerts and notifications. For simplicity and ease of understanding, the case study was generated from the perspective of a researcher in a healthcare company that has many patients with diabetes, which requires both ongoing treatment and regular health monitoring. As seen in figure \ref{Figure:DiabetesTreatmentandMonitoringDFD}, it is thus necessary to gather and analyse data from a Continuous Glucose Monitor (CGM) sensor device worn by patients. This sensor measures glucose levels constantly, taking readings at consistent intervals across several days. A researcher can thus use an application that can detect a set blood glucose level or ongoing change in such levels as a trigger. This application must analyse the gathered data and produce a notification to both the patient and the required professionals as well as any researcher. This would require any relevant health professional to have access to patient data for follow-up purposes, such as to allow the patient to be instructed to adjust their insulin dosage, do exercise, or change medication.

\subsubsection{Applicable Privacy Patterns}

Among the available Privacy Preserving measurements \cite{Alkhariji2020}, the most suitable measurements for software developer to use are the privacy patterns. For this specific use case, there were 55 out of 74 applicable privacy patterns. The next step was thus to allocate the appropriate privacy patterns and extract the CQs relevant to the use case. 

\subsubsection{Allocating Privacy Patterns among the DFD nodes}

\begin{figure*}
	\centering
	\includegraphics[scale=0.45]{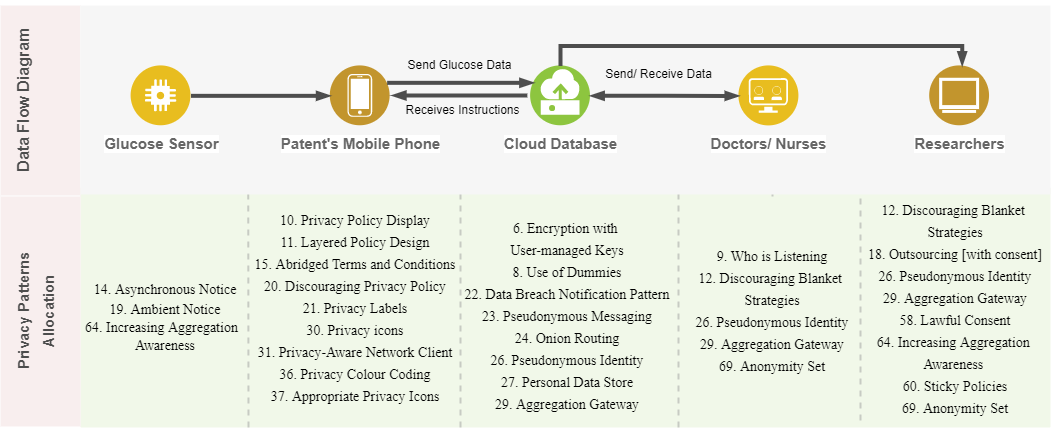}
	\caption{ {Privacy Patterns allocation of Health Care use case. It shows a list of Privacy Patterns for each node in the DFD. It also shows the Privacy Patterns that should be applied across all DFD nodes.}}
    \label{Figure:DiabetesTreatmentandMonitoringPrivacyPatterns}
\end{figure*}

Based on the researchers' knowledge and experience, the applicable privacy patterns were allocated among the DFD nodes of the use case, as shown in figure \ref{Figure:DiabetesTreatmentandMonitoringPrivacyPatterns}. Each node in the DFD operates some type of information or data activity, thus ensuring that there is a set of associated privacy patterns that propose privacy-preserving practices appropriate to that node. For example, storing information in the cloud entails the application of the privacy patterns: \textit{8. Use of Dummies} and \textit{63. Added-noise measurement obfuscation}, among others. Some privacy patterns are applied across multiple DFD nodes, however, and these were listed separately for ease. An example of this is the Privacy Pattern \textit{24.Onion Routing}, which refers to encrypting information during transfer, as this pattern should be applied across all DFD nodes. This step helped with developing a practical approach to identifying the CQs that might be asked based on each individual node and its related privacy patterns.

\subsubsection{Determining CQs based on the nodes in the DFD}
From the use case, 12 specific CQs were developed, each formulated based on the relevant DFD nodes and their related privacy patterns. For example, the \textit{researchers} node, which must considered to be a third-party factor, inspired the CQ \textit{What are the PbD patterns I should apply if my system shares data subject information with a third party/another organisation?} Some nodes might inspire more than one CQ in this way: for example, the \textit{mobile phone} node has many capabilities, thus inspiring the CQ \textit{What PbD patterns should I apply if my system includes a mobile phone?}, which can only be answered by the provision of a set of privacy policy display patterns. The same node also inspired the CQ \textit{What are the PbD patterns I should apply if my system requires the collection of personal data?}, which is answered by a set of privacy patterns related to the concepts Aggregate, Minimise, and Obfuscate. 

\subsection{CQs Developed within Workshops}
\begin{figure*}
	\centering
	\includegraphics[scale=0.55]{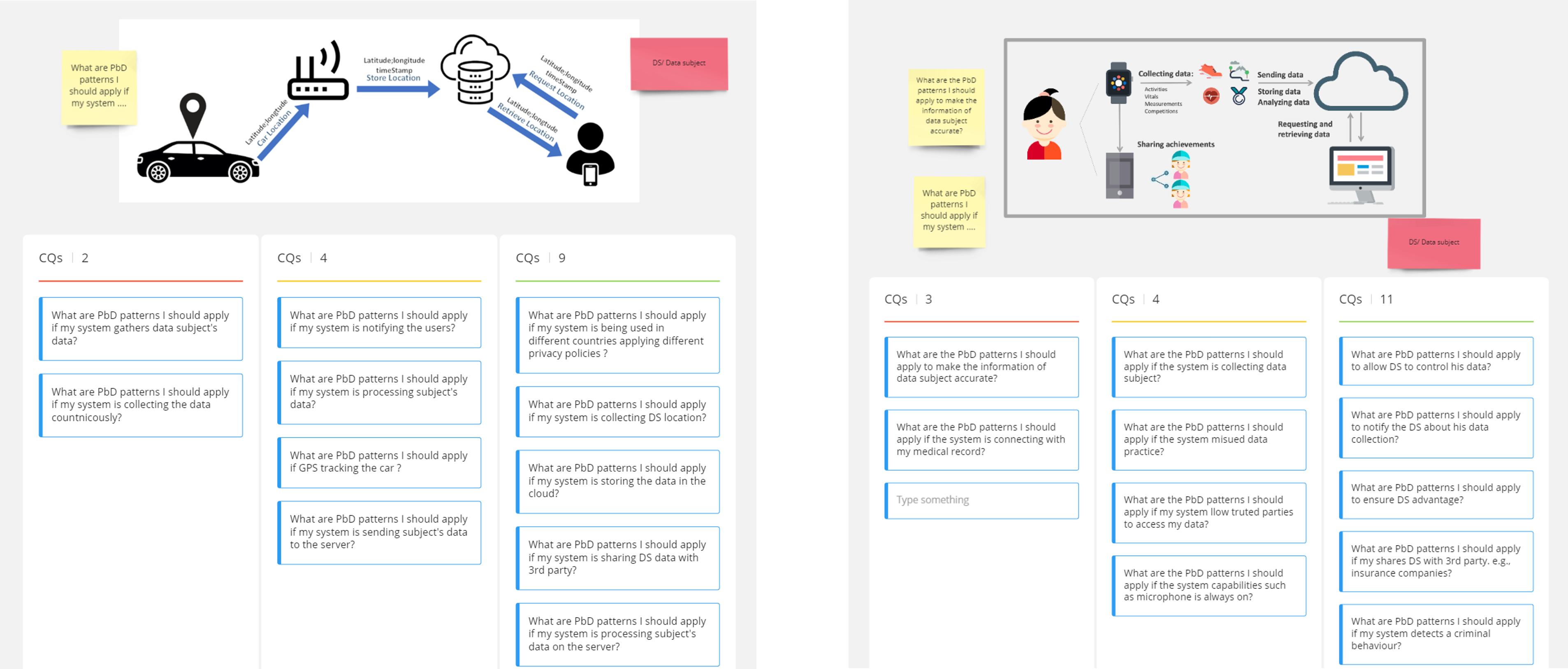}
	\caption{Screenshot of the conducted online workshops to gather CQs that are related to a given use case DFD. The screenshot shows two use cases and the CQs found by participants. This figure is provided in expanded form in \cite{orca149337}.}
    \label{Figure:workshops}
\end{figure*}

Three online workshops were held between May and June 2021 with various groups of software developers with diverse experience and knowledge about privacy-preserving measurements and practices, and work experience ranging from novice to expert. The workshops each began with a short presentation explaining the aim of the workshop to the participants and telling them what was required from them. Each workshop presented two use cases and explanation, with participants then given 25 minutes to discuss and write CQs for each use case. The underlying question in the workshops was: \textit{Based on the given IoT use case's DFD, what questions would you ask a privacy expert if you needed to apply relevant privacy-preserving measurements?}. Each workshop took an hour, with two participants involved in each session. Zoom software was used to facilitate the necessary online meetings, with Miro, an online collaborative whiteboard tool, used to share the use cases with participants in a manner that allowed them to add their CQs and notes freely. In this paper, we present the CQs as we got them from the participants, they might have some grammar and spelling issues. 
Figure \ref{Figure:workshops} shows two screenshots of the online workshops.

Across all workshops, five use cases were presented, all of them drawn from the IoT field. After showing the DFD for each use case, the main functionalities of the system were explained to participants.


These diverse examples offer a wide overview of the types of information that needs to be modelled in the PARROT ontology. The use case descriptions and DFDs can also be found in \cite{orca149337}.

The workshops generated 170 CQs across all the use cases; each CQ was then ranked depending on its usability. Seven statuses emerged: Valid, Duplicated, Modified, Discarded (out of scope), Discarded (No privacy pattern applicable), Discarded (No Existing privacy pattern), and Discarded (Duplicated within a use case).

\begin{itemize}[leftmargin=1em] 

 \item \textbf{Valid:} Valid ranked CQs were those related to relevant PbD aspects and answerable by means of applicable privacy patterns. For example, the Fitness Watch use case included the Valid CQ \textit{What are the PbD patterns I should apply if my system stores a user's daily routines?} A total of 30 Valid CQs was thus identified.

\item \textbf{Duplicated:}
Duplicated ranked CQs were those that addressed the same issue across different use cases. These CQs were considered separately, as each use case has a different list of applicable privacy patterns, depending on the scenario, leading to potentially different answers for each use case. An example might be the CQs \textit{What are the PbD patterns I should apply if my system is storing data in the cloud?} This question is duplicated in both the fitness watch system and the real tracking location system. In the fitness watch system, we can apply the privacy pattern \textit{65.  Attribute Based Credentials} but not in the real tracking location system. This is because the user needs the exact location of the car. We have a total of 31 Duplicated CQs.

\item \textbf{Modified:} 
Modified CQs were otherwise valid CQs where participants had formulated their questions in an unclear or an indirect format; these CQ were thus modified for clarity. An example emerged in the park monitoring system, where the question \textit{What PbD patterns should I apply if my system does not make people aware that a camera is recording them?} to \textit{What PbD patterns should I apply to make people aware that the camera is recording them?} A total of 19 Modified CQs thus emerged.

\item \textbf{Discarded (Out of Scope):}
Out of scope CQs were those not related to PbD aspects, which were thus out of scope of the current research.  For instance, in the Drone Delivery system case, the question \textit{What privacy patterns I should apply if my system is physically damaged?} emerged. The development case could not hold this type of question thus, it was discarded. A total of 23 Out of Scope CQs was identified.

\item \textbf{Discarded (No privacy pattern applicable):}
No privacy pattern applicable CQs are valid CQs where the use case does not attract any applicable privacy patterns to handle the stated issue. This rank of CQs mainly appeared in the Park Monitoring use case due to the limited numbers of applicable privacy patterns (20 privacy patterns). An example of such a question was \textit{What are the PbD patterns I should apply if my system is attacked while sending data from the router to the cloud?} The privacy pattern that could best handle this issue is \textit{22.  Data Breach Notification Pattern}, which is not applicable to the given use case, as it is intended only for use in cases that provide an interface with the data subject. This status was given to 8 CQs overall.

\item \textbf{Discarded (No existing privacy pattern):}
No Existing privacy pattern CQs were valid CQs where, due to a lack of available privacy patterns, no existing privacy patterns could handle the stated issue. Example of this included \textit{What are the PbD patterns I should apply if my system gathers interaction and behavioral data?} and \textit{What are the PbD patterns I should apply if my system gathers biometric data?} These CQs were discarded, though this does highlight the need to introduce new privacy patterns, which is discussed further in the future work section, as this status was given to 19 CQs.

\item \textbf{Discarded (Duplicated within a use case):}
Duplicated within a use case CQs were valid CQs arising from different participants covering the same issues; one of these was thus designated a Duplicated CQ in each case. For example, in the Smart Home System case, both the CQs \textit{How do I deal with outdoor cameras if they record strangers?} and \textit{What PbD patterns should I apply if my system collects data that is highly dependent on the external environment?} were offered. This status was given to 39 CQs. 

\end{itemize}

\subsection{Answering the CQs}

After deleting and amending the necessary CQs, consideration were made on a total of 81 CQs arising from the workshops. The workshops generated a list of applicable privacy patterns for each use case, with each CQ then answered manually based on researcher experience with a list of privacy patterns applicable to the relevant use case. The analysis of these is explained in more detail in the Analysis section. 

To answer each CQ, all applicable privacy patterns were examined to determine whether they resolved the issue raised in CQ; if so, the relevant privacy pattern was added to the answer list. During this process, the categories stated for each privacy pattern were used as guidance in terms of making the correct decisions. For instance, where a CQ asked about storing data for a period of time, the relevant privacy patterns were identified under the Inform category.


\section{Analysis} \label{Analysis}

This analysis describes the topics and the categories of the CQs and their answers (i.e. privacy patterns list for each CQ). In this section, we will explain the method we used to analyze the information we have, the types of CQs, and the tags of the privacy patterns.


\subsection{Method}
Collating all CQs collected from both researchers and workshops that were ranked valid, duplicated, or modified could be answered with an appropriate set of privacy patterns generated a list of 81 CQs in total. To expose the differences in these CQs across different use cases, these were then organised into six groups, as shown in table \ref{CQs Groups}.

\subsection{Findings: CQ Types}

Within the full list of CQs, several targeted different aspects of the systems’ DFD nodes or descriptions, i.e., the devices used or data activity types. Based on this, the CQs were classified into four main types and eight sub-types to allow distinct groupings to emerge. The following describes these types and sub-types shown in table \ref{CQs types}.

\begin{table}
\footnotesize
\centering
\begin{tabular}{c  c} 
 \hline
 \textbf{Number} & \textbf{Use case} \\ 
 \hline 
 11 CQs & Health Care System  \\ 
 30 CQs & Fitness Watch  \\
 10 CQs & Real Tracking Location System (RTLS)  \\
 9 CQs & Park Monitoring System  \\
 10 CQs & Smart Home System  \\ 
 9 CQs & Drone Delivery System  \\
 \hline

\end{tabular}
 \caption{Use cases used to gather information needs, with the number of CQs inferred from each use case.}
\label{CQs Groups}
\end{table}

\subsection{Findings: CQ Types}

Within the full list of CQs, several targeted different aspects of the systems’ DFD nodes or descriptions, i.e., the devices used or data activity types. Based on this, the CQs were classified into four main types and eight sub-types to allow distinct groupings to emerge. The following describes these types and sub-types shown in table \ref{CQs types}.

\begin{itemize} [leftmargin=1em]
\item \textbf{Data Collection.} These CQs relate to the various kinds of data that need to be collected within a system. This includes four sub-types: \textbf{Location}, referencing a data subject’s current geographic location, usually collected via GPS. \textbf{Personal Information}, which references any information that is specific to the data subject such as their name, personal address, phone number, etc. \textbf{Routine}, which is data that needs to be collected continually that can then be used to determine the data subject’s habits. An example of the latter type of CQ might be \textit{What PbD patterns should I apply if my system stores a user's food intake information?} Lastly, \textbf{Photo}, for CQs that consider the storing of raw or manipulated images of the data subject. The latter sub-type could also be split further into similar sub-types such as audio or video; however, the CQ list developed for this work did not warrant this step. Across the Data Collection category, 22 CQs emerged, five in the Location sub-type, 12 in the Personal Information sub-type, three in the Routine sub-type, and two for Photo sub-type.

\item \textbf{Device.} These were CQs regarding the kind of device specified, leading to the generation of four sub-types: \textbf{Mobile Phone, Camera, Microphone,} and \textbf{Reading Sensor}, potentially covering any sensor that could be used in the system. Eight CQs emerged in this category, one for Mobile Phone, four for Camera, two for Microphone, and one for Reading Sensor.

\item \textbf{Process.} These CQs considered the processes applied to data subject information across five sub-types: these were \textbf{Share}, for CQs about systems that allow the data subject to share information with friends or other users in the system; \textbf{Access}, for CQs about providing access to other users with different roles in the system, such as supervisors;, \textbf{Third-Party}, for CQs that address any issues regarding sending data subjects’ information to third-parties; \textbf{Route}, for CQs about transferring data subject information across different nodes in the system; and \textbf{Profile}, for CQs that consider data subject profiles and login issues. In this category, 20 CQs emerged, three for the Share sub-type, four for the Access sub-type, six for both sub-types, Third-Party and Route, and one for the Profile sub-type.

\item \textbf{Storage.} The CQs were around storing data subject information and the means and time periods appropriate to this. The distinction between collecting and storing data subject information can be confusing; this section was thus specified as including CQs that targeted issues arising after the act of collecting the information. Two sub-types emerged: \textbf{Cloud}, for CQs that considered issues arising from systems storing information in the cloud; and \textbf{Local}, for CQs that considered issues arising where the system stores information locally. In this category, five CQs emerged, four for Cloud and one for Local.

\item \textbf{Dignity.} These are CQs that consider the regulations and procedures that ensure a data subject’s dignity and right to privacy. This category thus contains four sub-types, \textbf{Advantage}, which refers to CQs that consider a data subject’s benefit during system use, considering equivalent advantages among all system users; \textbf{Agreement}, referencing those CQs that consider issues around making agreements with the data subject and upholding them, thus covering concerns such as consent, disposing of data, and obligations to perform the actions stated in any privacy policy documents; \textbf{Notify}, which refers to CQs that consider sending alerts to data subjects about any updates regarding their information, such as collection or breaches of data; and \textbf{Control}, for CQs that consider the data subject control over the collection and manipulation of their information. In this category, 25 CQs emerged, one for Advantage, 10 for Agreement, eight for Notify, and six for Control.

\end{itemize}

\begin{table}[]
\footnotesize
\centering
\begin{tabular}{llll}
\hline
Type                             & Sub-Type             & CQs & Total               \\ \hline
\multirow{4}{*}{Data Collection} & Location             & 5             & \multirow{4}{*}{22} \\ 
                                 & Personal Information & 12            &                     \\ 
                                 & Routine              & 3             &                     \\
                                 & Photo              & 2             &                     \\ \hline                                 
\multirow{4}{*}{Device}          & Mobile Phone         & 1             & \multirow{4}{*}{8}  \\ 
                                 & Camera               & 4             &                     \\ 
                                 & Microphone           & 2             &                     \\ 
                                 & Reading Sensor       & 1             &                     \\ \hline
\multirow{5}{*}{Process}         & Share                & 3             & \multirow{5}{*}{20} \\ 
                                 & Access               & 4             &                     \\ 
                                 & Third-Party          & 6             &                     \\ 
                                 & Route                & 6             &                     \\ 
                                 & Profile              & 1             &                     \\ \hline
\multirow{2}{*}{Storage}         & Cloud                & 4             & \multirow{2}{*}{5}  \\ 
                                 & Local                & 2             &                     \\ \hline
\multirow{4}{*}{Dignity}         & Advantage            & 1             & \multirow{4}{*}{25} \\ 
                                 & Agreement            & 10            &                     \\ 
                                 & Notify               & 8             &                     \\ 
                                 & Control              & 6             &                     \\ \hline
\end{tabular}
\caption{Types and Sub-types of CQs with the number of the CQs for each Type and Sub-type}
\label{CQs types}
\end{table}

\subsection{Findings: Privacy Pattern tags}
As the goal of this research is to indicate the relevant privacy patterns for a given system's DFD to improve privacy protection for the relevant the data subjects, data sets of combined CQs and answers were created, with each CQ assigned a list of privacy patterns that handle the privacy issues raised in the CQ. To help develop understanding of the types of treatment needed to handle the privacy issues raised in the CQs, descriptive classifications for all privacy patterns in each answer list were developed; these are referred to as tags. 

Hoepman \cite{Hoepman2018}: (1)\textit{Minimise}, (2)\textit{Hide}, (3)\textit{Separate}, (4)\textit{Aggregate}, (5)\textit{Inform}, (6)\textit{Control}, (7)\textit{Enforce}, and (8)\textit{Demonstrate} develops the assignment of tags to the privacy patterns that was based on the eight Descriptive Strategies. In previous research, the authors have examined each privacy pattern in conjunction with each Strategy to determine the relevant connections: the allocated tags can thus be found within \cite{alkhariji2020examining}. By referring to that examination, this work assigned the type of treatment needed to each CQ, based on the assumption that each answer could have more than one tag. For example, to handle the CQ \textit{What are the PbD patterns I should apply if my system stores people's faces/photos ( identifiable info)?}, the specified privacy patterns were tagged \textit{Inform} and \textit{Control}. 

Table \ref{CQs tags} shows the individual tags with the number of CQs assigned to each tag. Overall, \textit{Inform} and \textit{Control} were the most frequently used strategies with 65 and 52 CQs assigned to each, respectively. \textit{Demonstrate} had 36 assigned CQs, \textit{Minimise} had 22 assigned CQs, \textit{Aggregate} had 23 assigned CQs, and \textit{Hide} had 21 assigned CQs. The least frequently used strategies were \textit{Separate} and \textit{Enforce}, which had 16 and 10 assigned CQs, respectively.

\begin{figure}
	\centering
	\includegraphics[scale=0.27]{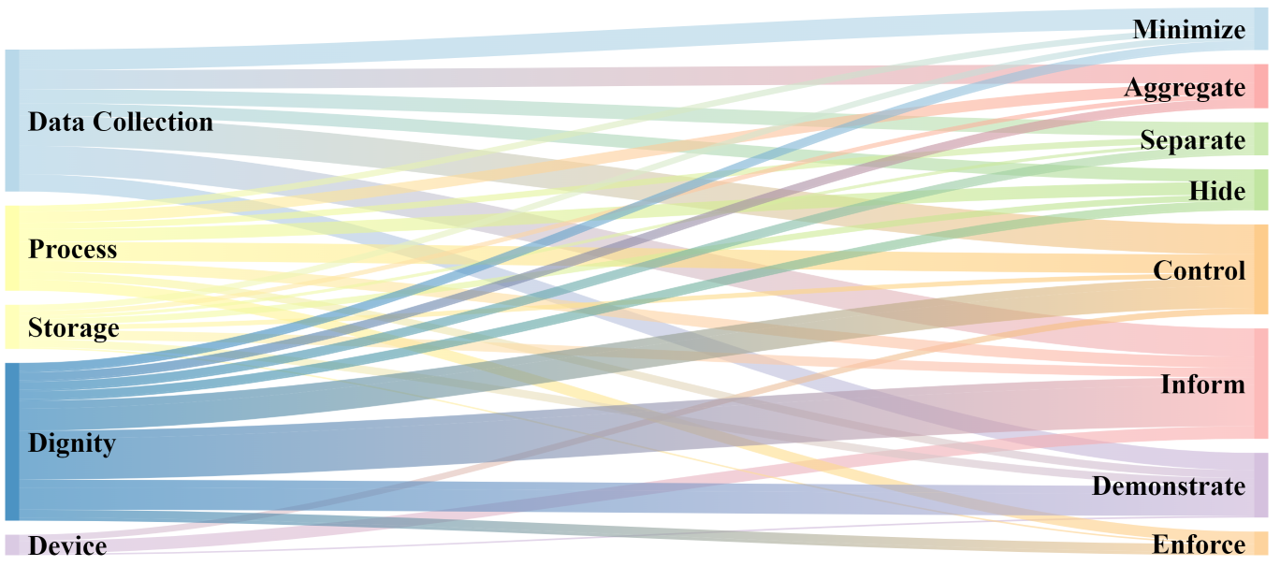}
	\caption{Diagram showing  analysis results. CQ types are shown on the right, while the left hand side highlights the tags given to the relevant answers.}
    \label{Figure:AnalysisSankeyDiagram}
\end{figure}

\begin{table}
\footnotesize
\centering
\begin{tabular}{c c} 
 \hline
 \textbf{Tag} & \textbf{Number of CQs} \\ 
 \hline 
 Minimize &  22 \\ 
 Hide  &  21 \\
 Separate &  16 \\
 Aggregate &  23 \\
 Inform & 65  \\ 
 Control & 52  \\
 Enforce &  10 \\ 
 Demonstrate &  36 \\
 \hline

\end{tabular}
\caption{Tags retrieved from Hoepman's eight strategies, with the number of CQs with privacy patterns answers assigned to each tag.}
\label{CQs tags}
\end{table}

 After analysing the data sets by assigning Types, sub-types, and tags, the answers to the CQs were reviewed within these groupings to ensure consistency between Types and tags for each CQ and its answer. For example, CQs of the Type \textit{Device} and sub-type \textit{Camera} would be expected to have similar tags to CQs under the Type \textit{Device} and sub-type \textit{Microphone}, as both CQs' answers should include the privacy patterns \textit{14. Asynchronous notice}, tagged as \textit{Inform}, and \textit{35. Enable/Disable Function}, tagged as \textit{Control}. After review, one CQ’s answer list was changed to Type \textit{Regulations}, and sub-type \textit{Control}. Initially, the CQ \textit{What are the PbD privacy patterns I should apply to allow data subjects to choose which of their data is collected?} was answered with privacy patterns tagged \textit{Minimise, Inform, Control}, and \textit{Demonstrate}; however, after reviewing CQs of the same sub-type, the \textit{Minimise} privacy patterns appeared irrelevant, and the privacy pattern \textit{2. Location Granularity}, tagged as \textit{Minimise} was removed, as this privacy pattern does not actually indicate allowing a data subject to choose the granularity level of a location, instead telling the controller not to collect an accurate location about the data subject. This privacy pattern was thus irrelevant to the answer list. Thus, all CQs of the \textit{Control} sub-type were tagged as \textit{Inform, Control,} and \textit{Demonstrate}. Diagram \ref{Figure:AnalysisSankeyDiagram} represents the flow between the types of CQs and their treatments. the colors in the diagram are auto-generated by Google charts \footnote{developers.google.com/chart/interactive/docs/gallery/sankey}, they do not represent any particular factor. The full list of CQ types and tags can be found in \cite{orca149337}.


\section{PARROT Ontology} \label{PARROT ontology}
In response to the need to develop an ontology that combines the knowledge held across the range of PbD explainable measurements and to apply this to IoT systems, the PARROT ontology was developed in the current work to cover a wider range of aspects. This section thus explains the PARROT ontology and how it was developed.

\subsection{PARROT Description}

\begin{figure}
	\centering
	\includegraphics[scale=0.7]{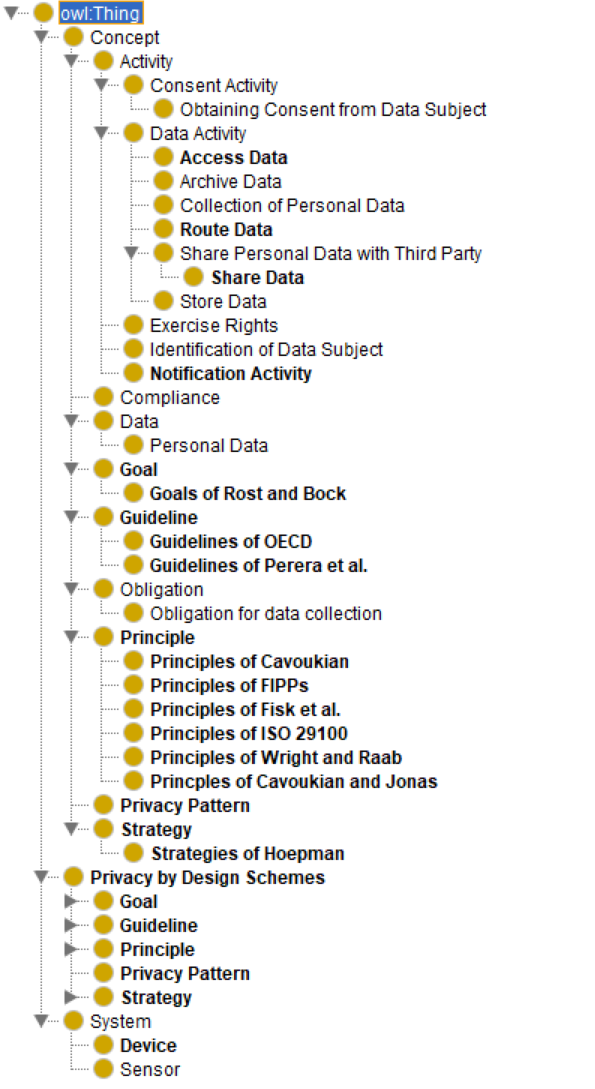}
	\caption{PARROT hierarchy, with the classes and subclasses used to model PARROT knowledge.}
    \label{Figure:PARROTHierarchy}
\end{figure}

OWL \cite{PascalHitzler} was selected as the initial ontology representation language, with PARROT then developed using Protégé 5.2.0, a popular open-source ontology editor. The PARROT ontology is currently formulated with 40 classes and subclasses, three object properties, and 239 individuals. The hierarchy of the PARROT ontology is shown in figure \ref{Figure:PARROTHierarchy}. Readers are referred to the supplementary technical report \cite{orca149337} for a more detailed explanation of ontology classes with their modeled object properties and instances and it is also available online\footnote{github.com/alkharijiLa/PARROT/blob/main/PARROT.owl}.
\newpage

\subsection{Searching Ontologies}
Although PbD is a new concept, an extensive search was made for ontologies that might model the required knowledge before developing a new ontology from scratch. Such PbD ontologies were searched for in two ways: (1) published papers and (2) ontology search engines. The reason for not depending only on the available ontology search engines is that these are not always updated to contain all recently published ontologies. For instance, on the date of searching, the search engine Linked Open Vocabularies (LOV)\footnote{https://lov.linkeddata.es/dataset/lov} showed dates in 2019 in the \textit{Latest insertion} section, suggesting some delay in updates.

Google Scholar was thus used to search for papers that developed or mentioned development of ontologies regarding PbD. The search query: "privacy by design" AND ("ontologies" OR "ontology") was input, which resulted in the identification of the 20 papers related to this research. After a review of each paper, 11 were found to be fully related to PbD, and from these, a list of 55 ontologies, including upper level, middle level, and specific domain ontologies, was made. On reviewing each ontology, only two ontologies emerged as suitable for reuse in the PARROT ontology: GDPRtEXT \cite{Pandit2018} and COPri \cite{Gharib2018}.

\subsection{Reused Ontologies}
To develop the PARROT ontology fully, the reuse of existing ontologies that model various concepts of the required knowledge was initially undertaken. Those ontologies were identified as outlined in this section.

\subsubsection{IoT Ontologies}
The aim of PARROT is to help to apply privacy patterns to IoT systems at the development stage; it was thus necessary to include IoT knowledge, including devices and specifications, in the ontology. A Semantic Sensor Network (SSN) ontology and the lightweight SOSA (Sensor, Observation, Sample, and Actuator) ontology were thus incorporated. These both describe sensors, actuators, and samplers as well as the resulting observations, actuation, and sampling activities \cite{Haller2017}.
From SSN, the class ssn:System and the relevant subclasses, such as sosa:Sensor, were used to model the instances of IoT devices in the data sets, such as the instance "Glucose Sensor".

\subsubsection{PbD Ontologies}
\setlength{\emergencystretch}{3em}
The GDPRtEXT ontology was included to support PbD knowledge, which describes the concepts defined, mentioned, and required by the General Data Protection Regulation (GDPR). GDPRtEXT uses the Simple Knowledge Organization System (SKOS) upper ontology, which provides a model for expressing the basic structure and content of concept schemes \cite{Pandit2018} \cite{Isaac}.
From the SKOS ontology, the class skos:Concept, along with many of its subclasses, was used to model the required knowledge of PbD. The class GDPRtEXT:Principle was also used to include all PbD principles, including the PARROT:Principles\textunderscore of\textunderscore ISO\textunderscore 29100. The class GDPRtEXT:PrivacybyDesign was also used to include all PbD schemes classes, such as PARROT:Strategies\textunderscore of\textunderscore Hoepman 

\subsection{Explainable PbD Knowledge}
The available PbD schemes were organised in levels (Principles, Strategies, Guidelines, Patterns) depending on their abstractness and specificity. They were then synthesised according to the relevant privacy patterns \cite{Alkhariji2020}. Privacy patterns are the most suitable level to describe the appropriate designs required for software engineers; however, to show them how these patterns adhere to more abstract PbD rules, full PbD scheme relationships were modelled in the PARROT ontology. 
All instances of PARROT:Privacy\textunderscore Patterns were linked with all instances of PbD schemes, including GDPRtEXT:Principle, PARROT:Strategy, PARROT:Guideline, and PARROT:Goal, via the object properties PARROT:fully\textunderscore inspired\textunderscore by and PARROT:partially\textunderscore inspired\textunderscore by. For example, the privacy pattern 'Abridged Terms and Conditions' was categorised as inspired by the various different PbD schemes.



\section{Validation by Testing the CQs} \label{Validation}

This section discusses the validation of the PARROT ontology by means of examination of the various Competency Questions (CQs). Ontology validation is that segment of ontology evaluation concerned with checking whether the ontology fulfills the specifications required for its intended use \cite{fernandez1997methontology} \cite{Vrandecic2013}. According to many ontology engineering methodologies, CQs are a form of ontology requirement specification, in that they reflect the needs that must be satisfied by the ontology \cite{Gomez-Perez} \cite{Gruninger1995}. 

\begin{figure}
	\centering
	\includegraphics[scale=0.36]{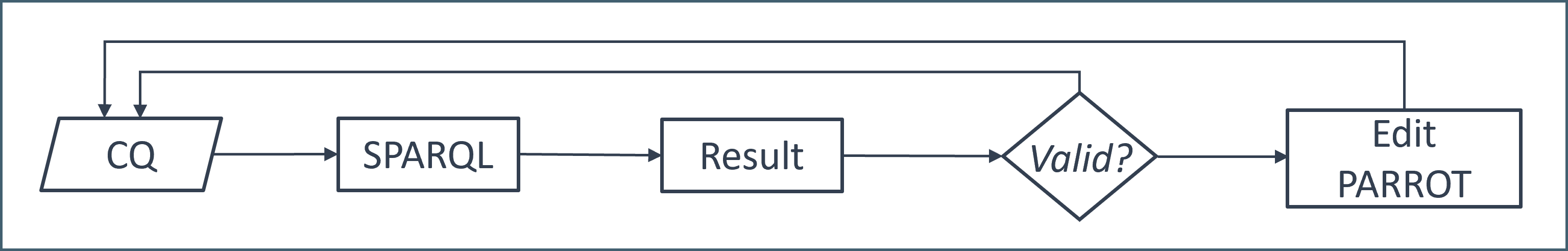}
	\caption{The validation process of the CQs begins with converting the natural language CQ into a SPARQL query, then running the query and checking the results. If the results are valid, the next CQ is run, and if not, the PARROT ontology is edited as appropriate.}
    \label{Figure:validation}
\end{figure}

In this research, therefore, the CQs were applied as validation criteria using SPARQL queries. The 88 CQs collected from different sources for the various use cases explained in section \ref{Gathering PARROT Information Needs} were applied, and, using Protégé, the informal CQs written in natural language were modified into formal CQs using SPARQL query \cite{gruninger1995role}. These queries were executed and checked for valid results. Where results are invalid in such cases, the ontology requires further improvements, as shown in figure \ref{Figure:validation}.

Table \ref{tab:Validation} shows a snapshot of the CQs, grouped in use cases, with the relevant SPARQL queries to provide the desired results. The full table can be found in \cite{orca149337}. The table descriptions are grouped by use cases and the classes were modeled in (Activity, Data, Device, Compliance, and Obligation). The CQs are also clustered as modelled in the subclasses of the PARROT ontology. 

\textbf{CQs 1-11} were collected for the health care system. CQ1 and CQ6 cover the \textit{devices} used in the system, a mobile phone and a glucose sensor, while CQ2, CQ3, CQ4, CQ5, CQ7, CQ8, CQ9, and CQ10 cover the \textit{activities} within the system: storing information, reporting for administration, sharing information with third parties, collecting personal data, routing data among system components, login functionality and providing data subjects with the ability to control their own data. CQ11 covers the \textit{data} to be collected in the system, including collecting the location of the data subject.

\textbf{CQs 12-21} were collected for the drone delivery system. In this use case, the CQs cover both \textit{activity} and \textit{data} to be collected in the system. CQ12 is about tracking activity, while CQ13 and CQ15 reflect notification activities in the case of a system attack or data infiltration. CQ14 and CQ20 are about retaining historical data in the archive of the system, while CQ16 is about providing the data subject with the ability to delete data. CQ18 is about giving access to data subject information to the service provider to enable the provider to carry out a specific service. CQ19 references 24-hour monitoring, while CQ21 is about collecting sensitive personal information, which is here considered to be a data concept, rather than an activity concept.

\textbf{CQs 22-52} were collected for the fitness watch system. These CQs covered the \textit{activity}, \textit{data}, and \textit{obligation} concepts. The \textit{activity} CQs were as follows: CQ22 and CQ43 are about exercising the right to ensure the data subject’s advantage and hiding the data subject’s identity from attackers. CQ24 is about the activity of collecting personal data where the device is always on, while CQ25, CQ38, CQ47, and CQ52 reference the activity of sharing personal data with a third party, selling data or sharing it with emergency services in case of an accident or where the data subject sends data to another user such as a friend. CQ26 similarly references about the activity of giving data access to trusted parties. CQ27, CQ39, and CQ42 are about the activity of collecting personal data anonymously, while CQ28, CQ30, and CQ49 are about notification activities in the cases of data collection or system attacks. CQ29 and CQ45 are about providing the data subject with the ability to control or delete their own data, CQ31 is about sharing data with another user, and CQ32, CQ33, and CQ36 are about storing data in the cloud or on a local device. CQ34 is about collecting real-time locations, while CQ35 is about storing the location of the data subject, and CQ37 is about keeping such data stored for some time. 

The \textit{data} CQs were as follows: CQ40 and CQ50 are about food intake or daily routine data, while CQ41 is about gym data. CQ44 and CQ46 are about city and address data, with CQ48 being about indoor positioning data. CQ51 is about data subject’s device properties data. CQ23 alone covers the \textit{obligation} concept, focusing on ensuring that the collected information is in line with the provisions stated in the privacy policy.

\textbf{CQs 53-61} were collected for the park monitoring system. These CQs covered both \textit{activity} and \textit{data} concepts. 
The \textit{activity} CQs were as follows: CQ54 is about providing data access to the administrator of the system, while CQ55 is about the controller’s compliance with respect to not giving access to non-authorised people. CQ58 is about routing data to the cloud, while CQ59 is about providing the data subject with the ability to delete their own data, and CQ60 is about storing data for some time. CQ61 is about whether the system has a live video monitoring functionality.
The \textit{data} CQs were CQ53 and CQ57, which are about facial image data, age and gender data, while CQ56 concerns the collected raw data.

\newpage
\begin{table}
    \centering
    \footnotesize{
    \begin{tabular}{p{1.5em} p{25em}}
    \toprule
    \multicolumn{2}{c}{\textbf{Health Care System}} \\
    \midrule
    CQ1.  & \multicolumn{1}{p{25em}}{What are the PbD patterns I should apply if my system includes a mobile phone?} \\
\cmidrule{2-2}          & SELECT ?Device ?PrivacyPattern\newline{}WHERE {  \newline{}?Device rdf:type PARROT:Device.\newline{}?Device PARROT:entails ?PrivacyPattern. \newline{}filter (?Device = PARROT:Mobile\_Phone ) } \\
    \midrule
    CQ2.  & \multicolumn{1}{p{25em}}{What are the PbD patterns I should apply if my system stores data subject information in a cloud-based database?} \\
\cmidrule{2-2}          & SELECT ?DataActivity ?PrivacyPattern\newline{}	WHERE { ?DataActivity a gdprtext:DataActivity.\newline{}		 ?DataActivity PARROT:entails ?PrivacyPattern.\newline{}		FILTER (?DataActivity = PARROT:Store\_Data) } \\
    \midrule
    CQ3.  & \multicolumn{1}{p{25em}}{What are the PbD patterns I should apply if my system reports data subject information to an Administrator/Controller?} \\
\cmidrule{2-2}          & SELECT ?Activity ?PrivacyPattern\newline{}	WHERE { ?Activity a gdprtext:SystematicMonitoring.\newline{}		 ?Activity PARROT:entails ?PrivacyPattern.\newline{}		FILTER (?Activity = PARROT:Report\_for\_Adminstration) } \\
    \midrule
    \multicolumn{2}{c}{\textbf{Drone Delivery System}} \\
    \midrule
    CQ12. & \multicolumn{1}{p{25em}}{What are the PbD patterns I should apply if my system provides tracking service?} \\
\cmidrule{2-2}          & SELECT ?Activity  ?PrivacyPattern\newline{}WHERE { ?Activity a gdprtext:CollectionOfPersonalData.\newline{}?Activity PARROT:entails ?PrivacyPattern.\newline{}FILTER (?Activity = PARROT:Tracking) } \\
    \midrule
    CQ13. & \multicolumn{1}{p{25em}}{What are the PbD patterns I should apply if my system was attacked?} \\
\cmidrule{2-2}          & SELECT ?Activity  ?PrivacyPattern\newline{}WHERE { ?Activity a PARROT:Notification\_Activity.\newline{}?Activity PARROT:entails ?PrivacyPattern.\newline{}FILTER (?Activity = PARROT:Notify\_System\_Attack) } \\
    \bottomrule

\end{tabular}%
}
    \caption{PARROT ontology validation via SPARQL queries. The table shows CQs grouped by use cases with the equivalent SPARQLs.}
\label{tab:Validation}%
\end{table}

\textbf{CQs 62-71} were collected for a real-time tracking system (RTLS). The CQs here were about \textit{activity} and \textit{data} concepts. With regard to the \textit{activity} concept, CQ62 is about storing data in the cloud, CQ63 is about sharing the data with a third party, and CQ64 is about whether the system is used between different countries with different privacy policies. CQ65 is about the activity of notifying the data subject, while CQ66 is about the activity of collecting data continuously. CQ67 and CQ71 are both about routing data to the server or between system devices, CQ68 is about the activity of processing data, and CQ69 is about the activity of tracking the data subject. Within the \textit{data} concept, CQ70 references accessing email and phone number data.

\textbf{CQs 72-81} were collected for the smart home system. These CQs cover both \textit{activity} and \textit{device} concepts. The \textit{activity} CQs were as follows: CQ72 is about sharing personal data with a third party, CQ73 is about providing the data subject with the ability to choose which data is collected or shared, and CQ74 is about routing data between system devices. CQ75 and CQ80 are about notifying the data subject in case of infiltration or intruders, whereas CQ76 is about exercising the right to store data securely held by the controller, and CQ78 is about obtaining consent from the data subject.  
The \textit{device} CQs are covered by CQ77, which is about the microphone device, and CQ79 and CQ81, which reference the outdoor camera and camera devices.

It is important to note that the CQs collected for various stated use cases were also modeled to answer the same question for any other relevant use case. For example, if a use case has a microphone component, the microphone was modelled in the ontology regardless of the other components in the use case; thus, any use case with a microphone component could be addressed using the PARROT ontology. This is applied to all duplicated CQs among different use cases, such as the many systems that store personal data in the cloud. In such cases, the common privacy patterns were modelled between the use cases for the duplicated CQs as noted in section \ref{Analysis}.



\section{Technology Evaluation} \label{Technology Evaluation}

In the ontology engineering domain, various options are available for validating and evaluating ontologies. In this research, the set of pitfalls published by Poveda Villalón et al. \cite{Poveda-Villalon2010} \cite{poveda2016ontology} were used; this involves 41 such pitfalls, classified over three dimensions \cite{Poveda-Villalon2010} \cite{Gangemi2006}: (1) the Structural dimension, which focuses on syntax and formal semantics; (2) the Functional dimension, which focuses on the conceptualisation of the ontology for its intended use; and (3) the Usability-profiling dimension, which focuses on the ontology’s annotations to address the communications context. To assess the PARROT ontology with respect to these pitfalls, three methods were applied: (1) the Protégé HermiT reasoner, (2) Evaluation with an ontology pitfall scanner (OOPS!), and (3) Evaluation by lexical semantics experts. Table \ref{pitfalls table} shows the 41 pitfalls across the three dimensions and specifies how each pitfall was assessed. The full pitfalls catalog can be found at (http://oops.linkeddata.es/catalogue.jsp). A pitfall may be assessed in more than one way, and several pitfalls are duplicated among the three dimensions; thus, each pitfall is included only with respect to its first appearance. The pitfalls \textit{P4, P6, P8, P10, P11, P14, P15, P16, P17, P18, P21, P22, P23}, and \textit{P36} were assessed using the Protégé Hermit reasoner, while pitfalls \textit{P2, P3, P4, P5, P6, P7, P8, P10, P11, P12, P13, P19, P20, P21, P22, P24, P25, P26, P27, P28, P29, P30, P31, P32, P33, P34, P35, P36, P37, P38, P39, P40}, and \textit{P41} were assessed using OOPS! Pitfalls \textit{P1}, and \textit{P9} were then assessed by lexical-semantic experts. The following section discusses the three methods used to assess the pitfalls in more detail:

\begin{table*}[htbp]
  \centering
  \normalsize{
    \begin{tabular}{cp{8em}p{31em}rrr}
    
    \multicolumn{3}{p{20.5em}}{\textbf{Pitfalls}} & \begin{sideways}\textbf{Protégé}\end{sideways} & \begin{sideways}\textbf{OOPS!}\end{sideways} & \multicolumn{1}{p{1.085em}}{\begin{sideways}\textbf{Experts}\end{sideways}} \\
    \toprule

    \multicolumn{1}{c}{\multirow{30}[7]{*}{\begin{sideways}Structural Dimension\end{sideways}}} & \multirow{12}[1]{*}{Modelling Decisions} & P1. Creating polysemous elements &       &   {} &   {\checkmark} \\
          &   {} & P2. Creating synonyms as classes &       & \checkmark     & {\checkmark} \\
          &   {} & P3. Creating the relationship “is” instead of using “subclassOf”, “in-stanceOf” or “sameIndividual” (Creating the relationship "is" instead of using "rdfs:subClassOf", "rdf:type" or "owl:sameAs") &       & \checkmark     &  \\
          &   {} & P7. Merging different concepts in the same class &       & \checkmark     & {\checkmark} \\
          &   {} & P14. Misusing “allValuesFrom” &  {\checkmark} &   {} &  \\
          &   {} & P17. Specializing too much a hierarchy (Overspecializing a hierarchy) &  {\checkmark} &   {} &  {\checkmark} \\
          &   {} & P21. Using a miscellaneous class &  {\checkmark} & \checkmark     &  {\checkmark} \\
          &   {} & P23. Duplicating a datatype already provided by the implementation language &  {\checkmark} &   {} &  \\
          &   {} & P24. Using recursive definitions &       & \checkmark     &  \\
          &   {} & P25. Defining a relationship as inverse to itself &       & \checkmark     &  \\
          &   {} & P26. Defining inverse relationships for a symmetric one &       & \checkmark     &  \\
          &   {} & P33. Creating a property chain with just one property &       & \checkmark     &  \\
          \cmidrule{2-6}  & \multirow{10}[2]{*}{Wrong Inference} 
             {} & P5. Defining wrong inverse relationships &       & \checkmark     &  \\
          &   {} & P6. Including cycles in a class hierarchy &  {\checkmark} & \checkmark     &  \\
          &   {} & P15. Using "some not" in place of "not some" &  {\checkmark} &   {} &  \\
          &   {} & P18. Overspecializing the domain or range &  {\checkmark} &   {} &  \\
          &   {} & P19. Defining multiple domains or ranges in properties &       & \checkmark     &  \\
          &   {} & P27. Defining wrong equivalent properties &       & \checkmark     &  \\
          &   {} & P28. Defining wrong symmetric relationships &       & \checkmark     &  \\
          &   {} & P29. Defining wrong transitive relationships &       & \checkmark     &  \\
          &   {} & P31. Defining wrong equivalent classes &       & \checkmark     &  \\
\cmidrule{2-6}          & \multirow{5}[2]{*}{No Inference} & P11. Missing domain or range in properties &   {\checkmark} & \checkmark     &  \\
          &   {} & P12. Equivalent properties not explicitly declared &       & \checkmark     &  \\
          &   {} & P13. Inverse relationships not explicitly declared &       & \checkmark     &  \\
          &   {} & P16. Using a primitive class in place of a defined one &  {\checkmark} &   {} &  \\
          &   {} & P30. Equivalent classes not explicitly declared &       & \checkmark     &  \\
\cmidrule{2-6}          & \multirow{3}[2]{*}{Ontology Language} & P34. Untyped class &       & \checkmark     &  \\
          &   {} & P35. Untyped property &       & \checkmark     &  \\
          &   {} & P38. No OWL ontology declaration &       & \checkmark     &  \\
    \midrule
     \multirow{7}[6]{*}{\begin{sideways}Functional Dimension\end{sideways}} & {Real World  } & P4. Creating unconnected ontology elements &  {\checkmark} & \checkmark     &  \\
          &   {Modelling} & P10. Missing disjointness &  {\checkmark} & \checkmark     &  \\
\cmidrule{2-6}          & Requirements Completeness & P9. Missing domain information &       &   {} &  {\checkmark} \\
\cmidrule{2-6}          & \multirow{4}[2]{*}{Application context} & P36. URI contains file extension &  {\checkmark} &   {} &  \\
          &   {} & P37. Ontology not available on the Web &       & \checkmark     &  \\
          &   {} & P39. Ambiguous namespace &       & \checkmark     &  \\
          &   {} & P40. Namespace hijacking &       & \checkmark     &  \\
    \midrule
     \multirow{6}[6]{*}{\begin{sideways}Usability-Profiling \end{sideways}} & \multirow{2}[2]{*}{Ontology Clarity} & P8. Missing annotations &  {\checkmark} & \checkmark     &  \\
          &   {} & P22. Using different naming conventions in the ontology &  {\checkmark} & \checkmark     &  \\
\cmidrule{2-6}          & \multirow{1}[2]{*}{Ontology } & P20. Misusing ontology annotations &       & \checkmark     &  \\
          &   {Understanding} & P32. Several classes with the same label &       & \checkmark     &  \\
          &   {} & P37. Ontology not available on the Web &       & \checkmark     &  \\
\cmidrule{2-6}          & Ontology Metadata & P41. No license declared &       & \checkmark     &  \\
    \bottomrule
    \end{tabular}%
    }
    \caption{Table shows the classification of the common pitfalls and their used evaluation criteria }
  \label{pitfalls table}%
  \end{table*}%

\subsection{Protégé HermiT Reasoner}
HermiT\footnote{http://www.hermit-reasoner.com/} is a reasoner for ontologies written using the Web Ontology Language (OWL) that can be used to determine if an ontology has various necessary qualities such as consistency and satisfiability. In this evaluation, both Protégé and the HermiT reasoner were used to verify the PARROT ontology against selected pitfalls across three dimensions. With respect to the \textbf{structural dimension}, the PARROT ontology did not suffer from either the \textit{P14. Misusing “allValuesFrom”} and \textit{P17. Specializing too much (Overspecializing a hierarchy)} pitfalls. For pitfall, \textit{P21. Using a miscellaneous class}, most of the PARROT ontology’s classes were reused from previously verified ontologies, so they do not suffer from this; further, when the remaining names of classes were created, thus was influenced by the existing naming themes, preventing this issue.

HermiT did not detect any cases of \textit{P6. Including cycles in a class hierarchy}, nor \textit{P18. Overspecializing the domain or range}. The features in \textit{P15. Using "some not" in place of "not some"}, \textit{P16. Using a primitive class in place of a defined one}, or \textit{P23. Duplicating a datatype already provided by the implementation language}, were not used, so the PARROT ontology naturally does not have these pitfalls. With respect to the pitfall, \textit{P11. Missing domain or range in properties}, all domains and ranges of the PARROT ontology were fully defined. 

Within the \textbf{functional dimension}, HermiT detected the pitfall, \textit{P4. Creating unconnected ontology elements}, as object properties had mistakenly been added for some individuals without connection; this was thus fixed. HermiT also detected the pitfall, \textit{P10. Missing disjointness}, allowing the disjoint axiom to be removed from the classes \textit{PARROT:Principle}, \textit{PARROT:Guideline},  \textit{PARROT:Strategy}, and \textit{PARROT:Privacy\_Patterns}.

For the \textbf{usability-profiling dimension}, manual checking was applied to the annotations for all classes defined for the PARROT ontology, with any missing ones added to check \textit{P8.Missing annotations}. The names of classes and individuals were also manually checked to satisfy \textit{P22.Using different naming conventions in the ontology}.

\subsection{Ontology Pitfall Scanner (OOPS!) Evaluation}
OOPS! is a tool that helps ontology developers to evaluate ontologies automatically by detecting a subset of the most common pitfalls in ontology development \cite{Poveda-Villalon2012}. The tool interface provides many options prior to evaluation, and in this case,  the PARROT ontology’s RDF full text was pasted in the ontology box. This included all classes and properties imported from other ontologies (skos, ssn, sosa, and GDPRtEXT). 
The OOPS! scanner evaluates all ontology elements, including those from imported ontologies. However, only the PARROT ontology elements, which start with the “\textit{PARROT}” prefix were examined. The pitfalls were prioritised by labelling in three colours depending on the importance level of the pitfall (red: critical, orange: important, yellow: minor). A screenshot of the OOPS! scanner results and a table of the pitfalls found in the PARROT ontology for each dimension are offered in \cite{orca149337}.

\subsubsection{Structural Dimension }
The Structural dimension has four sub-classifications: (1) Modelling Decisions (2) Wrong Inferences (3) No Inferences and (4) Ontology language. A figure shows a screenshot of the evaluation results and a table of the pitfalls and their occurrences for the structural dimension in the PARROT ontology are laid out in \cite{orca149337}. This shows several pitfall occurrences; however, considering only the PARROT elements, nine occurrences require further examination. The minor pitfall \textit{P07.Merging different concepts in the same class} occurred in three cases,  \textit{PARROT:Principles\_of\_Wright\_and\_Raab , PARROT:Principles\_of\_Cavoukian\_and\_Jonas}, and \textit{PARROT:Goals\_of\_Ros t\_and\_Bock}. However, it is the “\textit{and}” phrase in these classes that causes the scanner to believe that these classes merge two concepts, while in fact they are one principle as listed by two researchers. The minor pitfall \textit{P13. Inverse relationships not explicitly declared} also occurs in three cases; however, in the PARROT ontology, the relationships created are not invertible. The critical pitfall \textit{P19. Defining multiple domains or ranges in properties} similarly occurs in three PARROT relationships, as the OWL language allows an object property to have more than one domain or range. This problem was addressed by typing all domains and ranges in the “Class expression editor” as a single entry. In this way, Protégé is enabled understand the domain and range as a union of or intersection of multiple classes to create a single domain/range.

\subsubsection{Functional Dimension }
\setlength{\emergencystretch}{5em} {
The functional dimension has three subclassifications: (1)Real World Modelling or Common Sense, (2)Requirement Completeness, (3)Application context. A screenshot of the evaluation results and a table of the pitfalls occurances are offered in \cite{orca149337}. Only one important pitfall appeared in this dimension, \textit{P10. Missing disjointness}. This pitfall applies to the ontology in general rather than any specific element, and to address this pitfall, some classes were specifically declared as disjoint classes. In particular, the subclasses of the \textit{GDPRtEXT:Principle} class, \textit{PARROT:Principles\_of\_Cavoukian}, \textit{PARROT:Principles\_of\_FIPPs}, \textit{parrot:Principles\_of\_Fisk\_et\_al}, \textit{PARROT:Principles\_of\_ISO\_29100}, \textit{PARROT:Principles \_of\_Wright\_and\_Raab}, and \textit{PARROT:Principles\_of\_Cavouk- ian\_and\_Joans} are all disjoint classes. The subclasses of \textit{PARROT:Guidline} class, \textit{PARROT:Guidelines\_of\_OECD}, and \textit{PARROT:Guidelines\_of\_Perera\_et\_al} were set as disjoint classes. 
}

\subsubsection{Usability-Profiling Dimension }

The usability-profiling dimension has three sub-classifi- cations: (1) Ontology Clarity, (2) Ontology Understanding, and (3) Ontology Metadata. A screenshot of the evaluation results and a table of the addressed pitfalls are given in \cite{orca149337}. The evaluation result showed 59 cases; however, when only the PARROT elements were examined, 12 cases presented the pitfalls that applied to the whole ontology. The pitfalls \textit{P07. Merging different concepts in the same class} and \textit{P13. Inverse relationships not explicitly declared} were addressed as described with respect to the first dimension, while the minor pitfall \textit{P08. Missing annotations}, which occurs in six PARROT classes was fixed by means of multiple comment annotations to the PARROT classes. These classes and annotations are shown in \cite{orca149337}. The minor pitfall \textit{P22. Using different naming conventions in the ontology} applies across all ontology elements, not to specific ones, as the reuse of multiple ontologies caused naming conventions to vary. In the PARROT ontology, an underscore is used between the words, with main words capitalised, as in the example: \textit{PARROT:Principles\_of\_Cavoukian}. Finally, to address minor pitfall \textit{P41. No license declared}, a license for the PARROT ontology will be declared in future work.

\subsection{Expert Evaluation}
A semantic web expert was then asked to evaluate the PARROT ontology against the pitfalls: \textit{P1. Creating polysemous elements}, \textit{P2. Creating synonyms as classes}, \textit{P7. Merging different concepts in the same class}, \textit{P9. Missing domain information}, \textit{P17. Specializing too much a hierarchy (Overspecializing a hierarchy)}, and \textit{P21. Using a miscellaneous class}. After learning about the ontology’s purpose and implementation, she provided feedback on the ontology by means of an evaluation form, found in \cite{orca149337}. The expert raised several issues, the most important of which are \textit{P1. Creating polysemous elements} and \textit{P17. Specializing too much a hierarchy}. All these issues were thus properly handled. With respect to pitfall \textit{P21}, she raised the issue that a class and an individual cannot have the same name, a constraint broken at that stage by $PARROT:Device$. The individual used as a miscellaneous element was thus removed, allowing retention of the class of that name. For the pitfall \textit{P17}, based on the expert's suggestion,the class $PARROT:Privacy\_by\_Design\_Schemes$ was moved to be a sibling of $skos:Concept$ class rather than being a subclass. Thus, $GDPRtEXT:Principle$ was no longer both a sibling and a subclass for $PARROT:Privacy\_by\_Design\_Schemes$.


\section{Content Evaluation} \label{Content Evaluation}

This section explains how the content of the PARROT ontology was evaluated by means of a user study. To achieve this evaluation, participants who had software developer titles were recruited, and the Wizard of Oz concept was applied to testing. Wizard of Oz is a methodology that allows testing of a prototype of a system before its actual development by means of having a person simulate the interface \cite{Dahlback1993}. The evolution study was thus structured as follows: (1) Each participant was provided with an IoT use case that included a description and a DFD diagram. (2) The participants were prompted to ask questions to a privacy expert in order to make the resulting systems privacy-preserving. (3) These questions were recorded and transcribed into an Excel file. (4) Their questions were then analysed and categorised them based on the categories built for the CQs, as explained in section \ref{Analysis}. (5) Checks were made to ensure that the PARROT ontology could answer any valid questions, and a list of the relevant answers was then extracted. (6) Finally, the software developers were provided with the privacy patterns lists that the ontology would suggest them to be applied in the system design. The following sections outline the approach taken for the pilot study and that applied to the rest of the participants.

Six use cases were distributed among 10 participants. They vary in their knowledge about privacy matters and their experience with software engineering. The use case descriptions were as listed in \cite{orca149337}. Each participant received two or three use cases randomly, which led to three participants examining the park monitoring system, and four participants each looking at the health care system, fitness watch system, smart home system, and drone delivery system. For the real-time tracking location system, five participants received the case. 

\subsection{Pilot Study}


A face-to-face pilot study was conducted with another software engineer and this took about 50 minutes, and the process was audio recorded. Initially, the participant was introduced to the idea of the research and her role, and she was then given two use case scenarios and asked to draw them both and ask questions about them. The first use case was the health care system, and the engineer took about 10 minutes to understand and draw the use case DFD using the PARROT tool. After that, she began asking questions from the DFD about how to preserve the data subject’s privacy. The second use case was the real-time tracking location system, and the engineer again took around 10 minutes to understand and draw the use case.

By the end of the study, the engineer was able to ask 31 questions, 19 with respect to the first use case and 12 with respect to the second use case. However, some questions were repeated across both use cases. After the session, the engineers notes were collected, and she was updated about the purposes of the experiment. She suggested supplying the DFD to participants instead of asking them to draw it in each use case, as this step took a long time, particularly in terms of adding more details for each node of the diagram. She also found some difficulty in terms of phrasing the questions she was supposed to ask, so she suggested providing sample or model questions for participants. Both suggestions were valuable, and thus carefully considered. The pilot also demonstrated that generating a text description of the use cases consumed excess participant time; the participants were thus offered detailed DFDs for the main study.

\subsection{Main Study}

\begin{figure}
	\centering
	\includegraphics[scale=0.25]{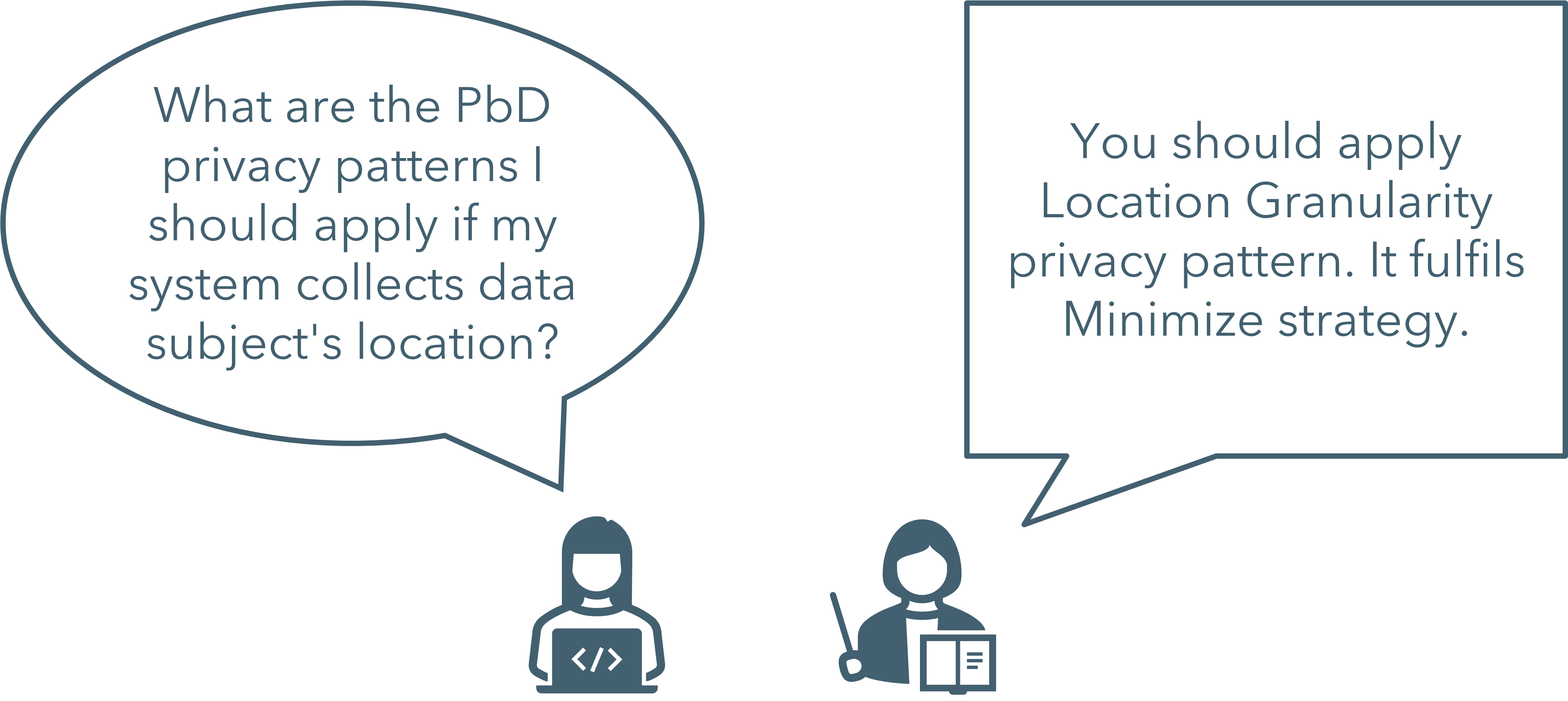}
	\caption{An example shown to participants before the beginning of the study, which shows a software developer asking a privacy expert a question about the system}
    \label{Figure:example}
\end{figure}

Based on learning from the pilot study, the participants in the main study were provided with the DFDs of the use cases rather than being asked to draw these. At the beginning of the session, the participants were assigned roles as software developers, and informed that the DFD was their outcome from initial specs. They were then asked to apply privacy-preserving measures to their systems, based on the fact that they personally did not have sufficient knowledge about what the necessary guidelines are and where to apply them to achieve this. They were thus informed that they had access to a privacy expert and that they could ask as many questions as they needed about applying the necessary privacy measures as illustrated in figure \ref{Figure:example}. 

Each participant was given two to three use cases, and sessions took between 15 and 30 minutes depending on the participant's understanding. 


\subsection{Study Analysis}
A total of 193 questions were extracted from participants. Before evaluating these questions, it was necessary to analyse and rank them as follows: 

\textbf{Valid}: The question is included within the ontology’s limitations. \textbf{Duplicated}: The question is valid, but has been asked more than once in some form. An example of this might be :\textit{ How long can I save the parking locations?} and \textit{For how long will the information be stored?}, which are variants of a question asked frequently across the study, sometimes repeatedly by the same participant across different use cases. \textbf{Discarded}: The question is invalid because it is out of scope or based on the software developer's decision. This occurred in some cases due to misunderstandings among participants based on them being unsure about what types of questions they should ask: thus, some questions were about how the system should work rather than privacy per se, such as "Does the system use face detection tools?" or "Is the cloud local or run by a third-party?" After filtration, 40 questions were discarded and 72 were identified as duplicated questions, leaving 81 valid questions. These valid questions were then assigned to the categories introduced in section \ref{Analysis}. 

The questions covered all types, though not all sub-types within those types. For the \textbf{data collection} type, 13 questions in total emerged, though for the \textbf{location} sub-type, no questions were offered. There were  10 questions for the \textbf{personal information} sub-type, and three questions for the \textbf{routine} sub-type. For \textbf{device} type, a total of 10 questions emerged, though the sub-type \textbf{mobile phone} received no questions, while the \textbf{camera} sub-type attracted two questions, the \textbf{microphone} sub-type got one question, and the \textbf{reading sensor} sub-type got seven questions. The \textbf{process} type got a total of 62 questions. Across sub-types, this divided into 11 questions for \textbf{share}, 18 questions for \textbf{access}, 12 questions for \textbf{third-party}, 16 questions for the \textbf{route}, and four questions for \textbf{profile}. The \textbf{storage} type similarly got a total of 29 questions, which, when divided into sub-types, represented 28 questions for \textbf{cloud}, and one question for the \textbf{local} sub-type. Finally, for \textbf{regulations} types, a total of 47 questions emerged. In terms of the relevant sub-types, this was three questions for the \textbf{privacy policy}, 32 questions for \textbf{agreement}, six questions for \textbf{notify}, and six questions for \textbf{control}. After sorting and classifying the collected questions, these were answered using the PARROT ontology.

\subsection{Results}
The final evaluation aimed to measure the extent to which the PARROT ontology can answer software developers’ questions about privacy-preserving measures in the design of IoT systems. In this step, each question ranked as \textbf{valid} was answered from the ontology by means of SPARQL queries. Across the 81 valid questions, this involved either the reused of the SPARQL queries from the validation list as mentioned in section \ref{Validation} or the creation of new SPARQL queries. However, the ontology struggled to answer all of the valid questions. 

Above and beyond the three ranks declared earlier (i.e., valid, duplicated, and discarded), two additional ranks for \textbf{valid} questions were thus created: \textbf{missing} and \textbf{not available}. A \textbf{missing} question is a valid question that is not yet covered in the ontology; such cases thus need to be modeled and added to the PARROT ontology. An example was \textit{Can I get the date of birth of the driver to check if he has a license?} A \textbf{Not available} question is also a valid question, but one for which there are insufficient privacy patterns to cover the issue raised. These questions should lead to the creation of a list of new privacy pattern suggestions for future work. Examples included \textit{If we have an external copy or backup of the information, how can I keep this private?} and \textit{If more than one person is using the watch, how do we protect all users' privacy?} Such questions require privacy patterns that are take into account the rights of multiple people, including those who do not use the service directly and who thus have not provided consent for their data to be collected. 

\begin{figure}
	\centering
	\includegraphics[scale=0.25]{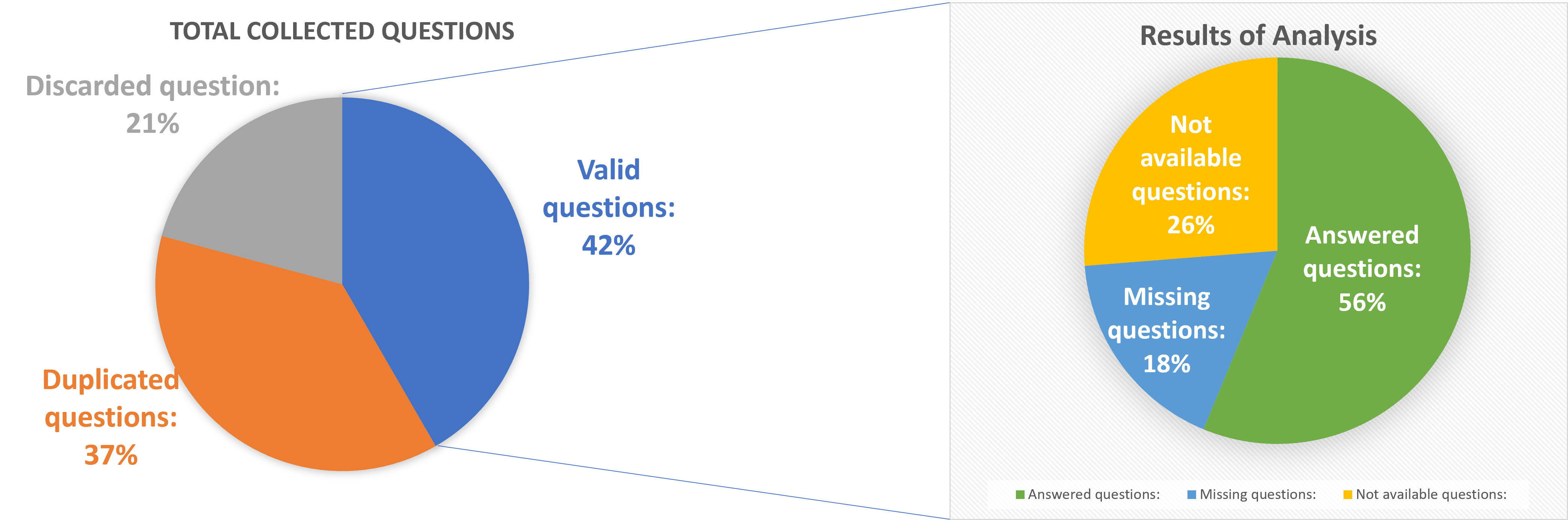}
	\caption{The results of the study analysis, showing that the PARROT ontology was able to answer 56\% of the questions. The ontology lacked the knowledge to answer 18\% of the questions, while 26\% of participant questions had no available answers.}
    \label{Figure:Results}
\end{figure}

Of the 81 valid questions, only 45 questions were answered successfully. Overall, there were 14 \textbf{missing} questions, and 21 \textbf{not available} questions, as shown in figure \ref{Figure:Results}. The full table of questions,  the accompanying analysis, and the relevant SPARQL queries can be found in \cite{orca149337}.

\section{Discussion} \label{Discussion}
The purpose of this project is to collect and formulate PbD knowledge and for application to IoT systems. This requires the development of an ontology that encapsulates the necessary knowledge to deliver answers to software engineers' questions during the design phase. This section thus discusses two relevant aspects of this research, the data set collection and formulation, the PARROT ontology representation, and the limitations of this research.

\subsection{Data Set}
Initially, to determine what sorts of potential questions software engineers might ask, three workshops were conducted with a range of participants examining six real IoT use cases. This led to the development of 81 CQs that were then modelled in the PARROT ontology that were assumed to reveal software engineers’ real needs to develop PbD measures and therefore to help improve the protection of data subject privacy in the resulting systems. These questions widely covered the issues in the use cases selected for this research, as well as covering some, but not all, issues arising in other use cases. Future work is thus necessary to expand the knowledge collected to adequately address additional IoT use cases.

\subsection{Ontology Representation}
Several existing ontologies, including some of the ones reused in this project, model the knowledge of IoT and other spheres with regard to privacy. However, to the best of the authors' knowledge, there is no existing ontology that models the relationships between IoT system components and PbD measures. In contrast, the newly developed PARROT ontology models not only the available PbD measurements, but also the relationships between them as a way to offer more explainable answer for software developers. 
The PARROT ontology was tested using the Wizard of Oz technique by 10 participants, and the ontology was able to answer 56\% of participant’s questions, a somewhat successful result. A possible explanation for this is that 26\% of the participant’s questions were not answerable by the PARROT ontology due to a lack of adequate relevant PbD measures that cover the issues raised in those questions. These questions should thus lead to the introduction of compulsory extended PbD measurements across the research community.

 \subsection{Scope and Limitations}

This research depends mainly on the mentioned six use cases. Further data collection based on extra diverse use cases is required to cover more privacy issues in the IoT systems. While the data set created in this research was developed by the researchers and participating software engineers, one limitation is that it could not address all the issues potentially arising in the stated use cases. This is because the knowledge modeled depended mainly on the questions raised while collecting the engineers' information needs. In addition, the content evaluation is conducted on 10 participants which offered a worthy examination of the ontology, yet, the evaluation needs to be made for not only a more comprehensive number of participants but also for a different set of use cases. This proposal provides a solution that is specified for IoT systems; however, it needs to be tested for a wider set of system fields to find if it is applicable for non-IoT environments.

Moreover, the available PbD measures cover multiple privacy threats and still overlook some aspects that need to be considered by the research community. Based on this, a new list of privacy patterns raised during the project journey is created, and the aim is to produce this in a separate work. Finally, The PARROT ontology does not offer the option for replacing or removing the system components, i.e., the ontology can not suggest an alternative for a camera device nor remove it if it is not necessary to provide the service, it instead will only indicate the proper guidelines for the camera component.

PbD measurements handle a wide range of aspects to preserve the data subjects' privacy, though, it has its own obstructions.  Looking into the strategies of Hoepman \cite{Hoepman2018}, we find that minimize and aggregate strategies cause a shortage of accuracy. Since this trade-off is subjective to the data subject, the control strategy allows the flexibility of prioritizing getting the full service with accurate data or saving some privacy for the data subject. This issue could also be addressed extensionally, as a future work, by modelling the PARROT ontology to comprise detailed explanation of the shortage of accuracy the PbD might cause so the software developer would be fully aware of the limitations and the options he provides for the data subjects in his design. 

\section{Conclusion and Future Work} \label{Conclusion}

The purpose of the current research was to determine whether the questions asked by software engineers about designing privacy-preserving IoT systems could be modeled and answered by a specific ontology. This research has thus identified the relevant knowledge that needs to be modeled appropriately to serve that purpose, and the PARROT ontology developed in this research thus contains two sectors of knowledge. The first is the Privacy by Design (PbD) measurements that were combined and analysed in previous work \cite{alkhariji2020examining}, while the second incorporates knowledge about IoT system devices and activities and how these should be addressed by such PbD measurements. This knowledge was collected by means of the development of Competency Questions (CQs) from six real IoT systems by researchers based on workshops with professional software engineers. This research also provided a categorised framework of the resulting questions and their treatments that lays groundwork for additional concerns about IoT systems to be addressed. 

The content of the PARROT ontology was assessed in a user study, which found that the PARROT ontology was able to answer 56\% of the questions asked. In general, the PARROT ontology encapsulates much of the available knowledge required to guide software engineers to apply PbD measures to their system designs, as well as explaining PbD measurements to software engineers by showing the connections to other measurements. It would also be advisable to link all PbD measurements to GDPR legislation to ensure that software engineers are fully aware of what laws they must comply with across their systems. Finally, the next step is to introduce of a chatbot interface for the PARROT ontology would allow software engineers to interact with and retrieve information more easily.


\bibliography{mybibfile}








\end{document}